\documentclass[12pt]{article} 

\begin{document}
\small{\it Eur. Phys. J. B}
\vskip 3.0cm

\begin{center}
{\Large \bf Domino effect for world market fluctuations}

\vskip 1.5cm 
{\large \bf N.Vandewalle, Ph.Boveroux and F.Brisbois}

\vskip 0.5cm
GRASP, Institut de Physique B5, Universit\'e de Li\`ege, \\ 
B-4000 Li\`ege, Belgium.

\end{center}

\vskip 3.0cm
{\noindent \large Abstract}
\vskip 0.6cm

In order to emphasize cross-correlations for fluctuations in major market places, series of up and down spins are built from financial data. Patterns frequencies are measured, and statistical tests performed. Strong cross-correlations are emphasized, proving that market moves are collective behaviors. 

\vskip 0.5cm
{\noindent Keywords: econophysics, critical phenomena } 

\newpage

Statistical physics started a few years ago to investigate financial data since they seem to exhibit complex behaviors, i.e. departures from true randomness. Various physical methods have been already reported to sort out correlations in financial data \cite{mantegna,dfa}. Recently, Bonanno et al. \cite{bonanno} studied data for 29 indices from different countries. The study demonstrated the existence of cross-correlations between these market places as well as a regional (continental) organization.

In order to emphasize and quantify the cross-correlations between the major financial indices around the world, we present here an analysis using a different approach. Our analysis distinguish up and down fluctuations. 

Figure 1 presents the closing values of three major financial indices from January 1980 till December 1999: the Japanese Nikkei, the German DAX and the Dow Jones Industrial Average (DJIA). Due to the earth rotation, the trading hours are of course different: from 9h00 till 15h00 (local time) in Tokyo, from 8h30 till 16h00 (local time) in Frankfurt, and from 9h30 till 16h00 (local time) in New York. Thus, there is only a small overlap during trading hours for the german DAX index and the DJIA. The considered period of 20 years corresponds to about $5200 \times 3$ data points. Below, only the sign of the daily fluctuations will be considered whatever its amplitude. 

Figure 2 illustrates the different sequences of spins that one can build from the three data series: (a) from the DJIA only and (b) from the three series together. Positive and negative fluctuations are represented by up and down spins respectively. During the whole period, a fraction of positive fluctuations is counted.  

First, let us consider each index evolution separately such as the DJIA. This evolution corresponds to the third vertical series of spins of Figure 2. For this series, a fraction $b=0.510$ of ``up" spins (a bias) is measured. In our analysis, only patterns of length 3 made of ``up" and ``down" spins, also called triplets, are considered from 3 consecutive trading days in NewYork. Thus, there exist $2^3=8$ possible different triplets. Since $b>\frac{1}{2}$, the most frequent pattern is expected to be the ``up-up-up" with a probability $f_e = b^3$ while the less frequent one is ``down-down-down" with a probability $f_e = (1-b)^3$. Those expected probabilities $f_e$ are illustrated in the histogram (in grey) of Figure 3. One should note that the counting of pattern frequencies is similar to the Zipf technique which was originally introduced in the context of natural languages \cite{zipf}. The Zipf analysis has been e.g. applied to correlated systems like DNA sequences \cite{dna} and also for investigating the distribution of incomes \cite{incomes}. The observed (measured) frequency of each pattern $f$ is reported in white in Figure 3. Error bars are indicated, and are calculated assuming a binomial distribution of spins taking the bias into account. No significant deviation from the biased random distribution (in grey) is observed in Figure 3. One concludes that correlations between the signs of daily fluctuations cannot be observed for the DJIA using this statistical analysis. Similar results have been obtained for the Nikkei and DAX indices. One should note that Zhang \cite{zhang} reported recently a similar statistical analysis on the New-York Stock Exchange (NYSE) index. He found correlations which may be associated with the bias not taken into account in his work. 

Consider now the three index evolutions of Figure 1 together, i.e. the spin series resulting from the lining of the three daily spins in a successive way as if the succession of spins is recorded around the world, as illustrated in Figure 2b. One should note that holidays do not take place at the same dates in different countries. These days containing any closed market are not considered in our measurements. Over all markets and for the whole 20 years period, a fraction $b=0.502$ of ``up" spins has been measured. Such a bias is negligible but will nevertheless be taken into account in the following discussion. The observed frequencies of triplets are plotted in Figure 4. Since $b=0.502$ close to $\frac{1}{2}$, the deviations from a uniform distribution is not visible in Figure 4. Error bars are indicated. Surprisingly, large deviations from the expected grey distribution are observed. The largest differences are observed for the ``down-down-down" and ``up-up-up" patterns. In these cases, the frequency is about $f=0.17$ instead of the $f_e=0.125$ expected for a random process, i.e. a relative difference of 44\%! These deviations from the grey distribution represents what is known as the ``domino effect" indicating that one place influences the next opening market. In particular, two negative (positive) fluctuations are usually followed by another down (up) fluctuation on the next market. In other words, major market places fluctuates in a cooperative fashion. This behavior seems to be symmetrical with respect to up and down patterns for the whole 20 years period. Except one work on price waves in french markets during the 19th century \cite{wave}, it is the first time to our knowledge that this Zipf-like method is applied to emphasize such correlations in between market places. 

One may ask if the strength of the domino effect is constant with time. It does not of course. Figure 5 presents the histogram for a period of 2 years preceeding the crash of 1987. During that period, there was some ``euphory" and the indices were growing at a high rate (about an annual return of 20\% for the DJIA), except for the DAX. The measured bias is thus quite large for that period: $b=0.538$. One observes also that the difference between the random and the observed distributions is quite large during that period with respect to the 20 years period investigated in Figure 4. In other words, stronger correlations are observed before crashes as suggested by recent works on the predictability of drastic events \cite{crash,sornette}. Another remark is that the differences between observed and expected frequencies for ``up-up-up" and ``down-down-down" triplets are not similar. Indeed, for the ``down-down-down" triplets, $f \approx 0.15$ instead of the $f_e \approx 0.10$, i.e. a relative difference of 50\% while for the ``up-up-up" triplets, $f \approx 0.20$ instead of the $f_e \approx 0.16$, i.e. a relative difference of 25\%. This result means that the correlations are more marked for ``down" spins than for ``up" spins.

Our analysis of up and down daily fluctuations is rather simple. One may ask for a more complicated analysis. We have recently shown \cite{zipfabud} that the use of other fluctuation types for describing for example large or small up and down fluctuations, i.e. four spin types, leads to other types of correlations and more visible structures. 

Statistical physicists love spin models because simple ingredients/rules make complex dynamics. Spins can represent up and down daily fluctuations. A daily fluctuations series as considered above can be viewed as the growth of a semi-open chain of successive up or down spins \cite{mem}. At each time step, a new spin is added at the extremity of the semi-open chain. Both histograms of Figures 4 and 5 mean that ``ferromagnetic" interactions have to be considered and that successive domains of up and down spins exists. Though the modelling ot the markets is outside the scope of the present paper, it suggests that modelling is possible in a physical (spin) framework like spin glasses \cite{mezard87}. Also, the physical quantities as the entropy, susceptibility or magnetization can be useful as market indicators for analysts.

In summary, we have performed some analysis for the daily evolution of three major world financial indices. It has been discovered that strong correlations exists between market places. Moreover, these correlations have been quantified such that the so-called domino effect is emphasized and quantified. It has been put also into evidence that the amplitude of the domino effect varies with time and seem to be more pronounced before a crash.

\vskip 1.0cm
{\noindent \large Acknowledgements}
\vskip 0.6cm

NV is grateful to the FNRS (Brussels, Belgium) for financial support. Valuables discussions with M.Ausloos, R.D'hulst, S.Galam, A.Pekalski, R.N.Mantegna, D.Stauffer and H.E.Stanley are acknowledged.

\newpage
{\noindent \large Figure Captions}
\vskip 0.6cm

Figure 1 --- Semi-log plot of three major world financial indices from January 1980 till December 1999: the Nikkei225, the DAX30 and the Dow Jones Industrial Average. Important financial events are emphasized.

\vskip 1.0cm
Figure 2 --- Typical examples of the construction of spins series from financial data series: (a) a single index and (b) three indices. 

\vskip 1.0cm
Figure 3 --- Histogram of triplets frequencies for the Dow Jones Industrial Average. Two cases are illustrated: the expected frequency from a random distribution taking the bias into account (in grey) and the observed frequencies (in white). Error bars are indicated. 

\vskip 1.0cm
Figure 4 --- Histogram of triplets frequencies for the lining series. Two cases are illustrated: the expected frequency from a random distribution taking the bias into account (in grey) and the observed frequencies (in white). Error bars are indicated.  

\vskip 1.0cm
Figure 5 --- Histogram of triplets frequencies for the two years period lining series preceeding the crash of October 1987. Two cases are illustrated: the expected frequency from a random distribution taking the bias into account (in grey) and the observed frequencies (in white). Error bars are indicated.

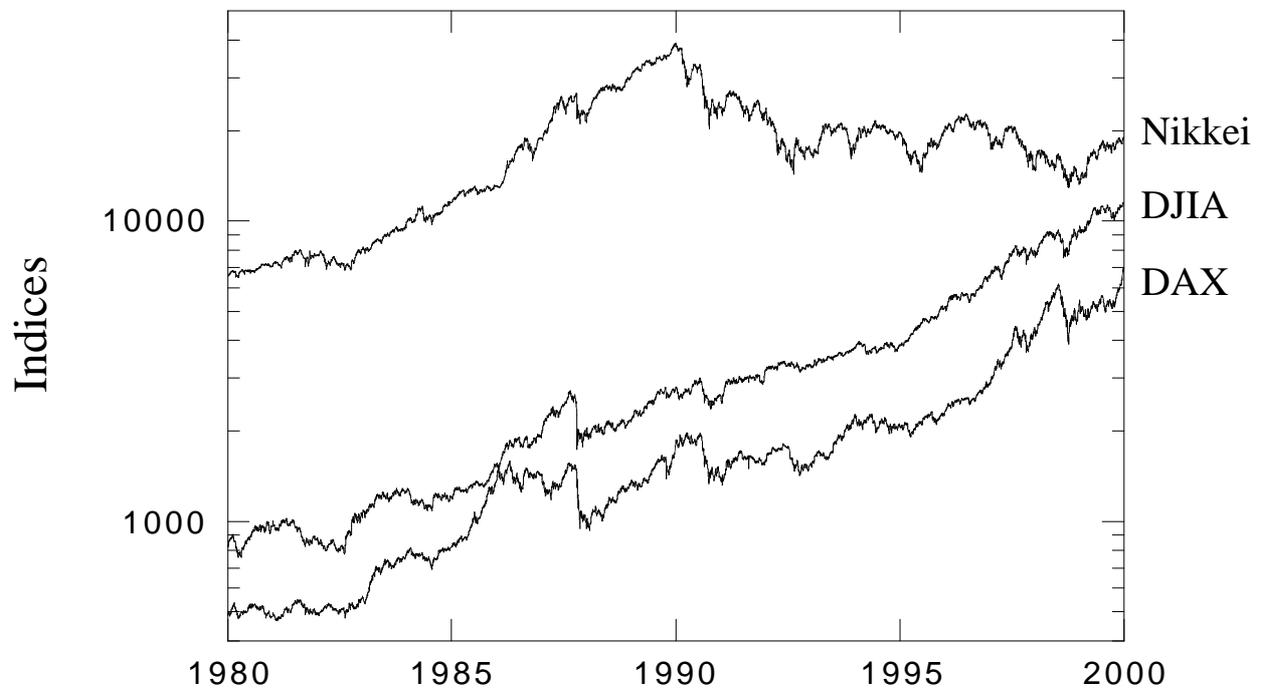
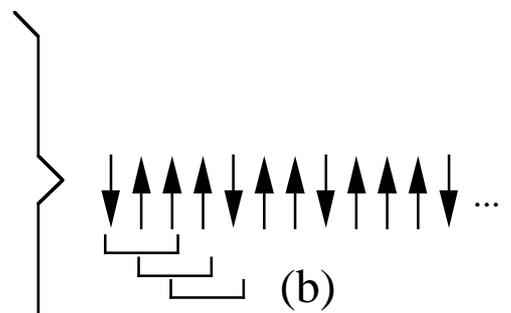
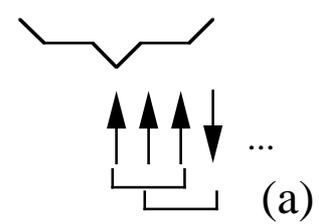
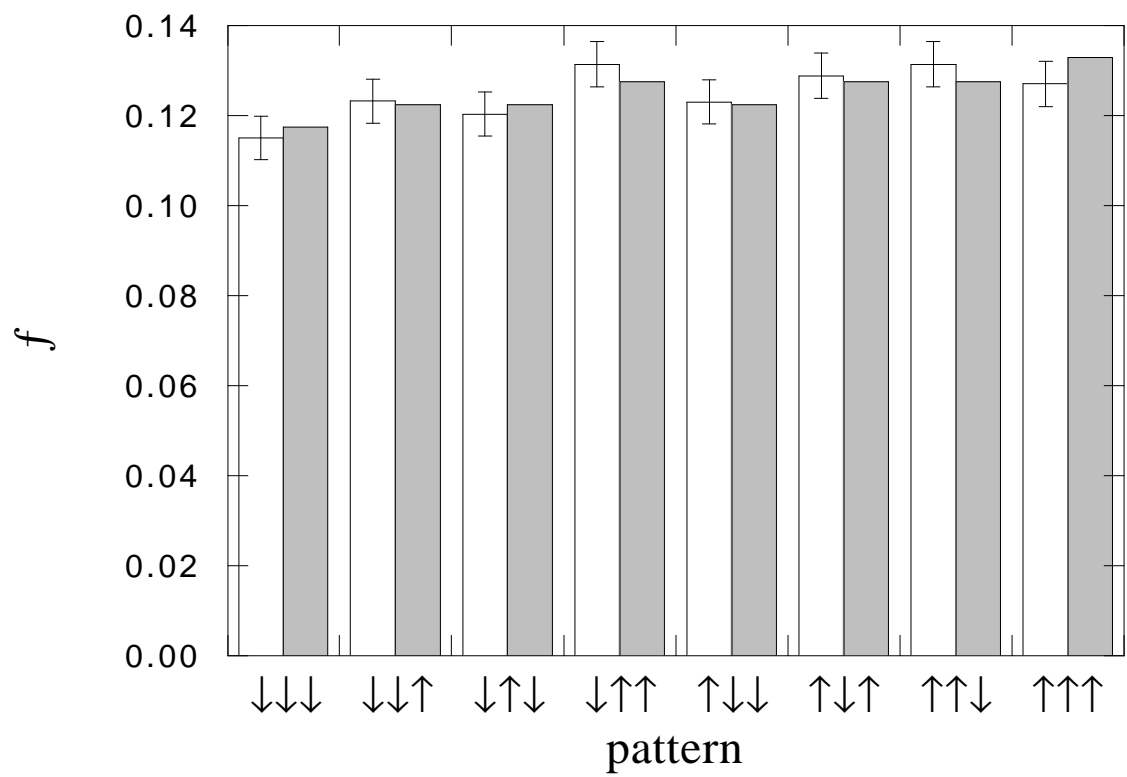
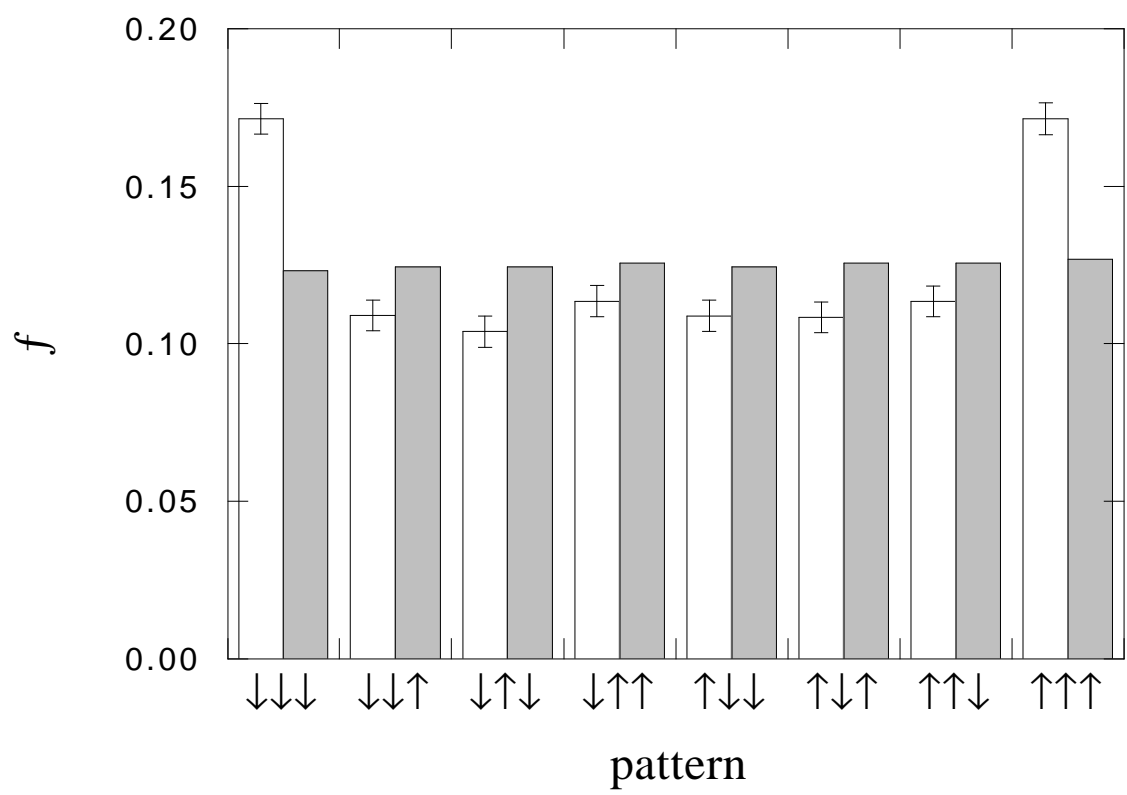
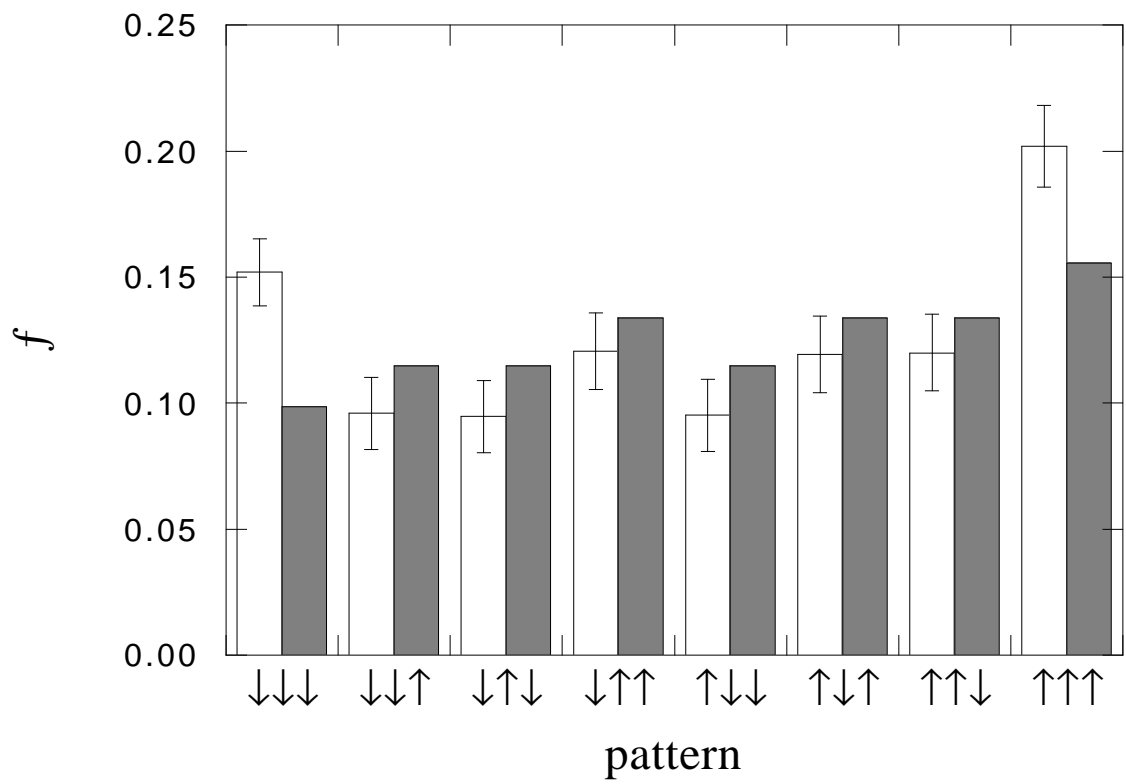
\end{document}